\documentclass[%
 reprint,
 amsmath,amssymb,
 aps,
]{revtex4-2}

\usepackage{graphicx}
\usepackage{dcolumn}
\usepackage{bm}
\DeclareUnicodeCharacter{0308}{{\"{}}}

\begin{document}

\preprint{APS/123-QED}

\title{Tunability of Robust Exciton-Trion Polaritons in Atomically Thin WS$_2$ Monolayers }

\author{Xuguang Cao$^1$, Debao Zhang$^1$, Ji Zhou$^1$, Wanggui Ye$^1$, Changcheng Zheng$^2$,\\ Kenji Watanabe$^3$, Takashi Taniguchi$^4$, Jiqiang Ning$^{1,*}$, and Shijie Xu$^{1,}$}

\altaffiliation{Corresponding author:\\
xusj@fudan.edu.cn\\jqning@fudan.edu.cn}

\affiliation{
$^1$Department of Optical Science and Engineering, School of Information Science and Technology, Fudan University, 2005 Songhu Road, Shanghai 200438, China\\$^2$Division of Natural and Applied Sciences, Duke Kunshan University, Kunshan 215316, China\\$^3$Research Center for Electronic and Optical Materials, National Institute for Materials Science, 1-1 Namiki, Tsukuba 305-0044, Japan\\$^4$Research Center for Materials Nanoarchitectonics, National Institute for Materials Science, 1-1 Namiki, Tsukuba 305-0044, Japan}%

\date{\today}

\begin{abstract}
Herein, we present an experimental demonstration of the robust exciton-trion polaritons (ETPs) by measuring and simulating the resonance reflectance spectra of various configurational WS$_2$ monolayers with different dielectric screenings. Moreover, the oscillator strength and decoherent behavior of such hybrid ETPs can be tuned via utilizing dielectric screening effect. The effect is attributed to the regulation of the Coulomb coupling between excitons and trions by changing the surrounding dielectric constant. The demonstration and tunability of the robust ETPs offers a novel pathway for researching novel phases of quantum matter in a quantum many-body physics regime.
\end{abstract}

\maketitle

{\section{\label{sec:level1}INTRODUCTION}}
The extraordinary atomic-layer transition metal dichalcogenide (TMDC) semiconductors represent a next-generation electronic and optoelectronic application candidate for device downscaling aspiration \cite{1, 2}. Unlike traditional semiconductor materials, a strong excitonic effect exists in these atomically thin systems largely due to reduced environmental dielectric screening, probably increasing excitonic binding energies $E_b$ to several hundreds of meV \cite{3}. In turn, the dielectric screening sensitivity offers the opportunity to further investigate novel electronic photophysical properties of these atomical layer materials, i.e., via observing the behaviors of the excitonic Coulomb interactions with changing dielectric-screening magnitude \cite{3, 4}. Various excitons and coupled electron-hole complexes, including strongly bound neutral excitons \cite{5}, charged excitons (trions) \cite{mak2013tightly}, and biexcitons \cite{you2015observation}, mainly originating from the direct optical bandgap transitions located at the finite momentum $K$/$K'$ points of the hexagonal Brillouin zone \cite{wang2018colloquium}, have been investigated in different two-dimensional (2D) TMDC semiconductors. The large oscillator strength of these excitons and bound excitonic complexes has allowed for enhancing light-matter coupling and the forming half-light half-matter exciton-polaritons \cite{liu2015strong}. And in single quantum well system, even without confined microcavity mode, the resonant structure in cryogenic reflectance spectrum is considered the result of the coupling between excited excitons and light due to the large excitonic oscillator strength, reflecting to the polariton effects \cite{berz1991exchange}. In most previous works, the focus has been put on the exciton-polaritons with single resonance excitations, such as neutral exciton- \cite{liu2015strong}, trion- \cite{emmanuele2020highly}, moiré exciton- \cite{zhang2021van}, and interlayer exciton-polaritons \cite{louca2023interspecies}. Specifically, recent reports have been experimentally proved that the called trions are Fermi polarons, new quasiparticles originated from optically created neutral excitons are coherently coupled to the electron-hole pairs of Fermi sea, in doped TMDC semiconductors, which should be described as a four-body Suris tetron picture rather than a three-body bound state \cite{sidler2017fermi, liu2021exciton, ni2025valley, efimkin2018exciton}. The four-body Suris tetrons (the following text still uses trion to refer to four-body Suris tetron), regarded as charge-neutral bosons, could exist in a coherent superposition with a bosonic photon, which explains the observation of trion-polaritons \cite{emmanuele2020highly, kyriienko2020nonlinear, rana2021exciton}. However, the exploration of hybrid elemental excitation interaction is remaining elusive. Recently, a coherent hybrid superposition of exciton, trion, and photon states, namely exciton-trion polaritons (ETPs), has been theoretically argued to exist in the electron-doped monolayer MoSe$_2$ with many-body theory \cite{rana2021exciton}. Such ETPs are calculated to exhibit three energy-momentum dispersion bands. Subsequently, a four-coupling state of plasmon-exciton-trion-charged biexcitons was observed in WS$_2$-silver nanocavities, exhibiting strong saturation nonlinearity \cite{wei2023charged}. The emergence of these fascinating coupling quasiparticles suggests a novel pathway for researching many-body physics and polaritonic device architectures.

\begin{figure*}
\includegraphics[width=0.8\textwidth]{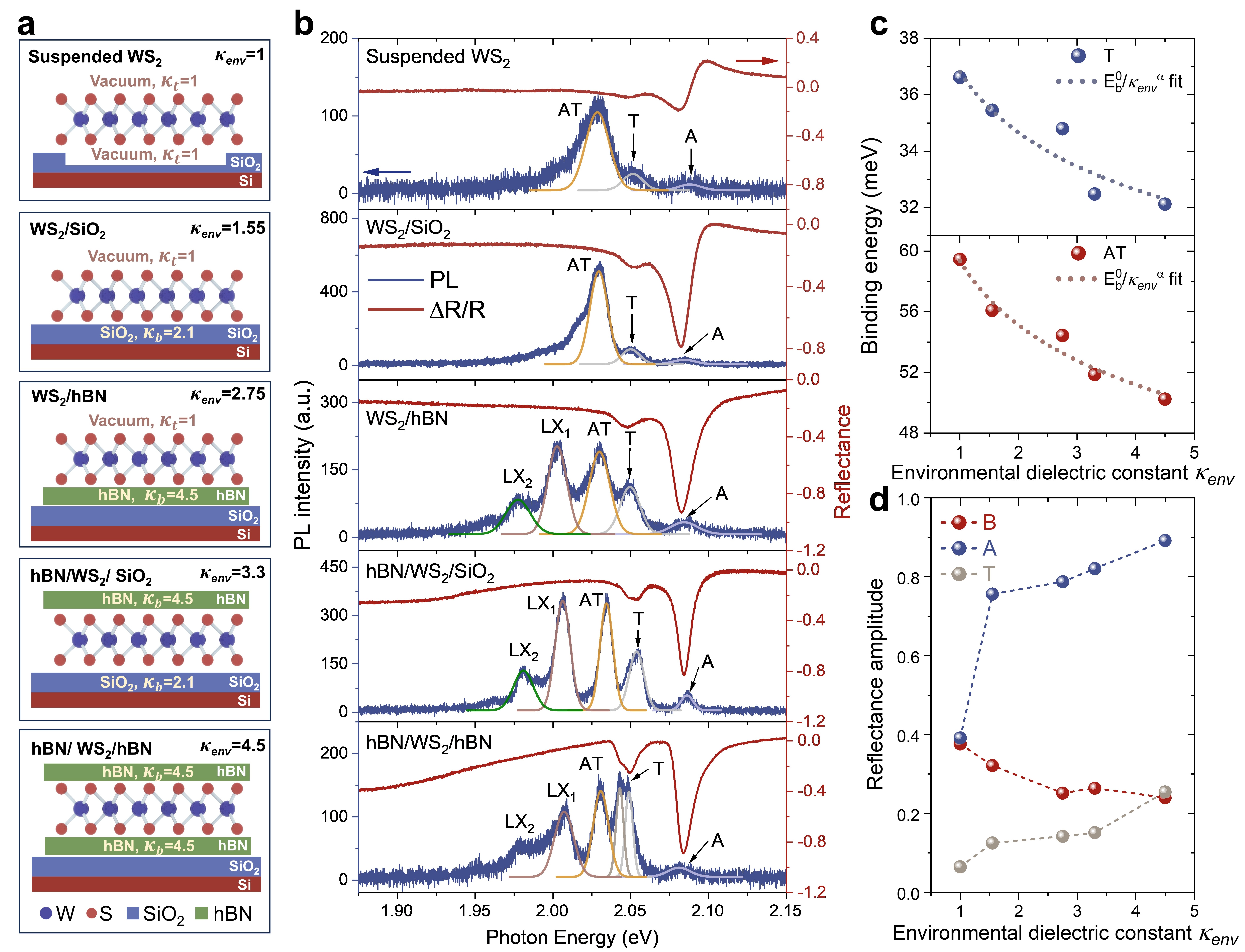}
\caption{\label{fig:wide}\textbf a Schematic diagram of the prepared samples with different dielectric screenings. Arranged from top to bottom are the suspended WS$_2$, WS$_2$/SiO$_2$, WS$_2$/hBN, hBN/WS$_2$/SiO$_2$ and hBN/WS$_2$/hBN samples, where the relative dielectric constants of SiO$_2$ and hBN are 2.1 \cite{3, 4} and 4.5 \cite{4, geick1966normal}, respectively. \textbf b Cryogenic PL (blue line) and contrast reflectance (brown line) spectra of the prepared samples measured under the identical conditions. Smooth solid lines with different colors represent the fitting curves to various exciton PL peaks with Gauss lineshape function \textbf c Dependence of the binding energies (solid circles) of the T trions and AT excitons on the surrounding dielectric constant. The dashed line represents a scaling relationship described by $E_{b(j)}$=$E^0_{b(j)}$($\kappa_{env}$)$^{(-\alpha_j)}$. \textbf d Dependence of the reflectance amplitude of the B, A and T excitons on the surrounding dielectric constant.}
\end{figure*}

In this work, we present a study of the dielectric responses of the composite ETPs in various monolayer WS$_2$ heterostructures by tuning the surrounding dielectric constant. Interestingly, the hybrid ETPs exhibit gradually enhanced oscillator strength with the increasing environmental dielectric screening. Furthermore, the specific influencing mechanisms of the environmental dielectric screening on the decoherence process of ETPs are unraveled. The enhanced dielectric screening mainly suppresses the recombination relaxation decay of the A exciton states in ETPs, including the relaxation to the dark exciton states via intervalley polariton-phonon scattering, whereas the pure decoherence rate of the trion states reduces. The attractively tunable behavior of ETPs is attributed to the Coulomb coupling enhancement, driven by approaching the resonance energies of the A excitons and the trions.

{\section{\label{sec:level2}EXPERIMENTAL METHODS}}
To tune the surrounding dielectric environment, five kinds of samples with different configurational architectures, namely suspended monolayer WS$_2$, monolayer WS$_2$ on SiO$_2$ (WS$_2$/SiO$_2$), monolayer WS$_2$ on hexagonal boron nitride (WS$_2$/hBN), monolayer WS$_2$ covered with hBN on SiO$_2$ (hBN/WS$_2$/SiO$_2$) and monolayer WS$_2$ covered with hBN on hBN (hBN/WS$_2$/hBN), were prepared by a dry transfer method. The schematic illustrations and optical images of the various sample structures are shown in Figure 1a and supplementary Figure S1, respectively. The monolayer WS$_2$ flakes were obtained through mechanical exfoliation of high-quality WS$_2$ bulk crystals, identified with optical microscopy and confirmed by both photoluminescence (PL) and Raman spectroscopies. The suspended WS$_2$ sample was fabricated by transferring a monolayer WS$_2$ flake onto a 10 $\mu$m-width groove etched on a SiO$_2$/Si substrate by reactive ion etching (RIE). The suspended part of the WS$_2$ flake over the SiO$_2$/Si groove is considered as in a freestanding state since the size of the suspended part (10 $\mu$m) is larger than the laser spot size (about 1 $\mu$m) used for optical spectroscopic testing. $\kappa_t$ and $\kappa_b$ are defined as the relative dielectric constants of the top covering material and the bottom supporting material, respectively (in a unit of $\kappa_0$, the vacuum permittivity). For the uncapped samples, the top surrounding environment is regarded as a vacuum state (testing pressure of 10$^{-4}$ torr, $\kappa_t$=1). Dielectric screening magnitude can be evaluated by the average dielectric constant of the surrounding environment, $\kappa_{env}$=($\kappa_b$+$\kappa_t$)/2. Among these samples, the suspended WS$_2$ sample has the weakest environmental dielectric screening ($\kappa_{env}$=1), and the hBN/WS$_2$/hBN sample experiences the strongest screening ($\kappa_{env}$=4.5) \cite{4, geick1966normal}.

{\section{\label{sec:level3}RESULTS AND DISCUSSIONS}}
\subsection{\label{sec:level3}DIELECTRIC SCREENING EFFECT}
The PL and reflectance spectral measurements were performed on a home-assembled multi-function-integrated micro-spectroscopic system, ensuring in-suit testing in all the samples. Detailed experimental setup is provided in the supporting information. Room-temperature PL spectra of the five testing samples obtained under the weak 532 nm laser excitation are shown in supplementary Figure S2a in the supporting document (unless otherwise specified, the testing laser power is 1 $\mu$W). One characteristic peak is observed and assigned to be the neutral A exciton resonance, confirmed by the contrast reflectance spectra (not shown). The cryogenic PL and the corresponding reflectance spectra (T= 6.5 K) measured under the identical conditions of the room-temperature spectra are depicted in Figure 1b. At the cryogenic temperature, the PL spectra comprise several peaks which are fitted very well with Gauss lineshape function. Among them is the emission from the neutral A excitons, a weak PL structure but with the highest energy, whereas its corresponding reflectance peak dominates the entire contrast reflectance spectrum since reflectivity is sensitive to the transitions of the free excitons with a large density of states \cite{cadiz2017excitonic}. The dominant reflectance structure of the A excitons indicates that low electron densities exist in these prepared samples. The PL peak energy of the A excitons exhibits a gradual reduction trend with the increasing $\kappa_{env}$ at low and room temperatures, which reflects that dielectric screenings indeed affect exciton behavior as the electronic bandgap is more sensitive to the dielectric environment compared to the excitonic binding energy \cite{3, xu2021creation} (see, for example, in supplementary Figure S2b). The succeeding emission feature (labeled as T) at lower energy is usually assigned to the emission of trions (i.e., as schematically shown in Figure 2a and upper panel of Figure 2b) \cite{mak2013tightly, rana2021exciton}. The dominant PL feature (labeled as AT) is attributed to the recombination emission of charged biexcitons, involving the binding of a bright neutral exciton to a dark trions \cite{you2015observation, nagler2018zeeman, chen2018coulomb}. And the following peaks, LX$_1$ and LX$_2$, may originate from phonon replicas emission of dark trions \cite{he2020valley,liu2019valley,li2019emerging}.

Based on the low-temperature PL spectra, the energy difference between the T (AT) structure and the A-exciton feature, that is, the binding energies of the T (AT) state, decreases with increasing $\kappa_{env}$, as shown in Figure 1c. Herein, it is necessary to carefully determine whether the reduction of the binding energies is primarily induced by the doping electron density or dielectric screening, where the former factor will significantly affect the binding energy due to the phase filling and free charge screening effects \cite{chernikov2015electrical}. We first figure out the relationship between the T/A integrated intensity ratio and the binding energies of the T and AT excitons from the cryogenic PL spectra of the several suspended WS$_2$ samples, to eliminate the environmental dielectric screening effects, where the T/A ratio has been shown to be positively correlated with the electron density \cite{mouri2013tunable}. It is found that the binding energy of trions exhibits a significant increasing trend with the increase of the T/A ratio in the several suspended WS$_2$ (supplementary Figure S3a), indicating the positive dependence of the trion binding energy on the Fermi energy in low electron density \cite{mak2013tightly,mouri2013tunable}. And the AT binding energy is insensitive to electron density, as shown in Supplementary Figure S3b. However, for the prepared samples with different environmental screenings, the extracted binding energies of the T and AT excitons both exhibit a negative dependence on their T/A ratio values. That is, the decreasing binding energies of the T and AT excitons discussed previously are primarily attributed to the increasing surrounding dielectric screening rather than the electron density. For the dependence of the excitonic binding energy on the surrounding environmental constant, a scaling relationship of $E_{b(j)}$=$E^0_{b(j)}$($\kappa_{env}$)$^{(-\alpha_j)}$, where $E^0_{b(j)}$ is the exciton binding energy in vacuum, $\alpha_j$ is an empirical scaling factor, \emph j index refers to the T and AT excitons, has been well established \cite{lin2014dielectric,kylanpaa2015binding}. Good consistency between experimental data and fitting curves (illustrated by dashed lines in Figure 1c) reveals that enhanced dielectric screening weakens the excitonic binding energies of T and AT states.

\begin{figure*}
\includegraphics[width=0.8\textwidth]{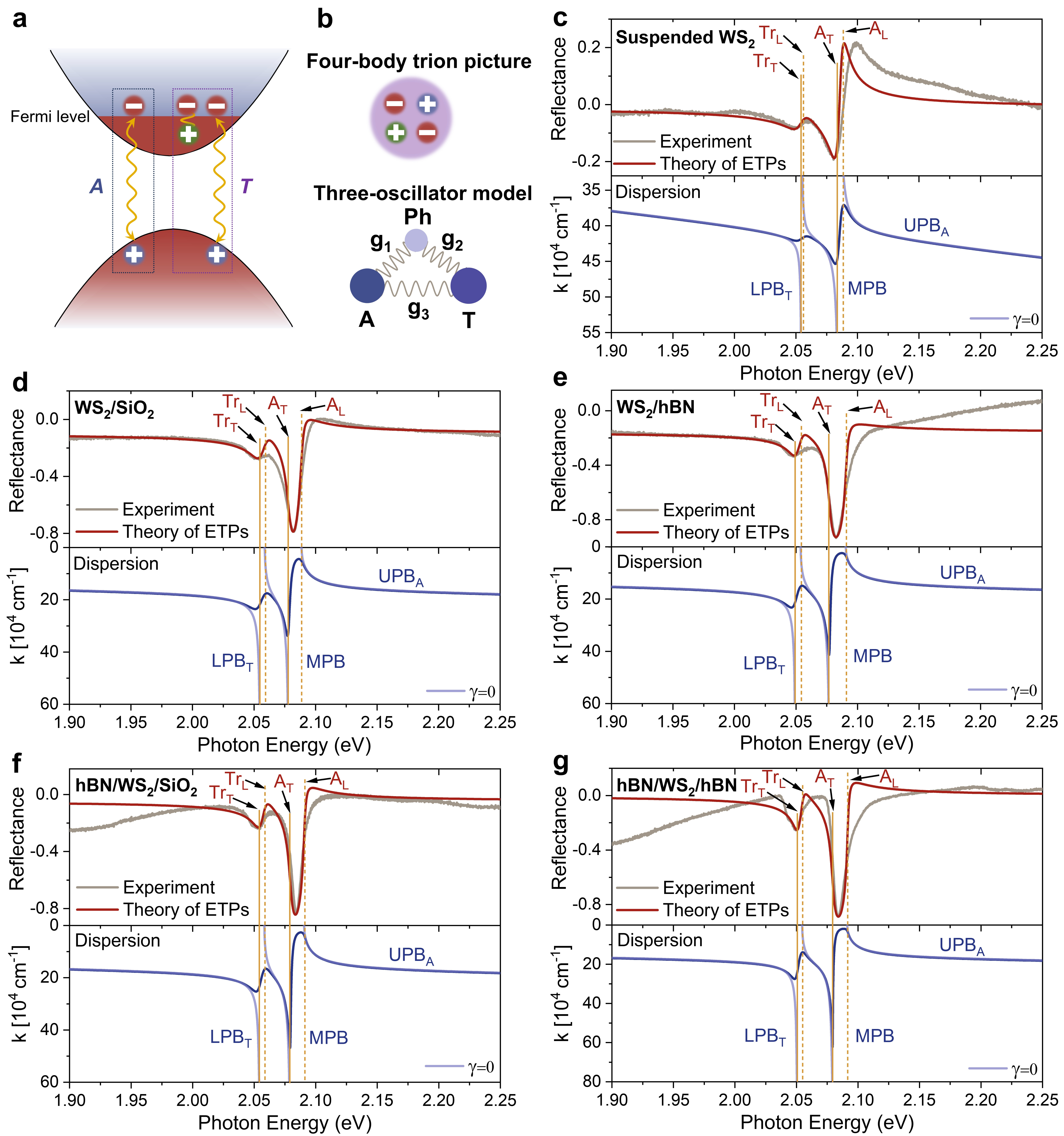}
\caption{\label{fig:wide}\textbf a Diagram of the conduction and valence bands of the monolayer WS$_2$ with the A, and T excitonic transitions. \textbf b Schematic of four-body trion (upper panel). And a coupled spring and mass system representing the three-oscillator model of photons (Ph), A and T excitons. g$_1$ and g$_2$ represent the exciton-photon coupling, and g$_3$ represent the Coulomb interactions between exciton and trion. Experimental and theoretical reflectance spectra (upper panel) and calculated polariton dispersion relation (lower panel) of the suspended WS$_2$ \textbf {(c)}, WS$_2$/SiO$_2$ \textbf {(d)}, WS$_2$/hBN \textbf {(e)}, hBN/WS$_2$/SiO$_2$ \textbf {(f)}, and hBN/WS$_2$/hBN \textbf {(g)} samples. }
\end{figure*}

\begin{table*}
\centering
\caption{\label{tab:table3}Values of various parameters adopted in the calculations of reflectance spectra of the ETPs in the series of monolayer WS$_2$ samples (Note: the unit of energy is “eV”).}
\begin{ruledtabular}
\begin{tabular}{cccccc}
 &suspended WS$_2$ & WS$_2$/SiO$_2$ & WS$_2$/hBN & hBN/WS$_2$/SiO$_2$ & hBN/WS$_2$/hBN \\ \hline
$\varepsilon_b$ & 15.4 & 2.7 & 2.3 & 2.8 & 2.8 \\
$\hbar \omega_{0_{(T)}}$ & 2.0546 & 2.0550 & 2.0496 & 2.0546 & 2.0506 \\
$\hbar \omega_{0_{(A)}}$ & 2.0850 & 2.0783 & 2.0774 & 2.0795 & 2.0795 \\
$4\pi \beta_T$ & 0.0055 & 0.0158 & 0.0150 & 0.0160 & 0.0170 \\
$4\pi \beta_A$ & 0.0210 & 0.0230 & 0.0250 & 0.0260 & 0.0279 \\
$\hbar \gamma_T$ & 0.0128 & 0.0112 & 0.0090 & 0.0084 & 0.0060 \\
$\hbar \gamma_A$ & 0.0070 & 0.0037 & 0.0026 & 0.0022 & 0.0017 \\
$\Delta_{LT_{(T)}}$ & 0.0004 & 0.0060 & 0.0067 & 0.0059 & 0.0062 \\
$\Delta_{LT_{(A)}}$ & 0.0014 & 0.0089 & 0.0113 & 0.0097 & 0.0103 \\
\end{tabular}
\end{ruledtabular}
\end{table*}

\subsection{\label{sec:level3}TWO-INTERACTING POLARITON CALCULATION}
The spectral resonances structures of the A and T excitons in reflectance spectrum are clearly observable, reflecting the resonance coupling of A and T state to their resonant photons. Considering the strong coupling between trions and photons, it is necessary to employ the four-body trions picture, as illustrated in Figure 2a that depicts a photo-generated exciton interaction with an electron-hole pair of Fermi sea. More interestingly, the A and T resonance structures show a notable dependence on environmental dielectric screening, with the resonant amplitude significantly enhancing as the $\kappa_{env}$ values increase (Figure 1d). The simultaneous increase in the reflectance amplitudes of A and T excitons indirectly confirms that the observed excitonic behaviors are not primarily induced by electron density, as it will cause the significant spectral weight shifts between the two excitonic resonances \cite{rana2021exciton,chernikov2015electrical}. The dielectric environment-dependent modulation of the reflectance spectral configuration is a highly intriguing phenomenon that necessitates a meticulous explanation. Considering the small energy separation between A and T state, the classical dielectric theory with two-interacting excitonic resonance were adopted to simulate the obtained experimental reflectance spectra to elucidate the underlying physical mechanisms \cite{berz1991exchange,lagois1977dielectric, schultheis1984reflectance}. The theory descripts the frequency versus wavevector polariton dispersion relation for two excitonic resonant structure with spatial dispersion, where the with two-interacting excitonic resonance are strongly coupled by Coulomb interactions \cite{lagois1977dielectric}. A schematic illustration of the three-oscillator model is shown in Figure 2b, where g$_1$ and g$_2$ represent the exciton-photon coupling, and g$_3$ represent the Coulomb interactions between A excitons and trions. The polariton property with two excitonic resonances showing spatial dispersion are given for a wavevector \emph k and frequency $\omega$ by a dielectric function $\varepsilon{(\omega,\emph k)}$ \cite{lagois1977dielectric}:
\begin{eqnarray}
\varepsilon{(\omega,\emph k)}=\varepsilon_b+\sum_{j}\frac{4\pi\beta_j\omega^2_{0(j)}}{\omega^2_{0(j)}-\omega^2+\frac{\hbar \emph k^2 \omega_{0(j)}}{M_j}+i\gamma_j \omega}
\end{eqnarray}

where $\varepsilon_b$ is the wavevector-independent background dielectric constant, \emph k is the wavevector, \emph j stands for the A exciton or the trion, $\omega_{0(j)}$ is the excitonic resonance frequency, $M_j$ is the total effective mass of the \emph j exciton, $\gamma_j$ is the wavevector-independent damping parameter, $4\pi\beta_j$ is the polarizability of the \emph j exciton, which could be described by introducing exciton-polariton oscillator strength $\emph f_j=4\pi\beta_j \hbar^2 \omega^2_{0(j)}$ (in a suit of eV$^2$). The transverse wave modes can be determined by the polariton equation\cite{lagois1977dielectric}:

\begin{eqnarray}
\varepsilon{(\omega,\emph k)}=\frac{k^2 c^2}{\omega^2}
\end{eqnarray}
where \emph c is the light speed in free space. Eq. (2) is cubic in $\emph k^2$ for A/T exciton-polariton calculation, leading to three transverse solutions for the coupled A and T excitons for a given principal polarization and frequency. The theoretical reflectance curves were numerically calculated by applying Maxwell’s electromagnetic boundary conditions and Pekar’s addition boundary condition \cite{pekar1958theory}.

\begin{figure*}
\includegraphics[width=0.8\textwidth]{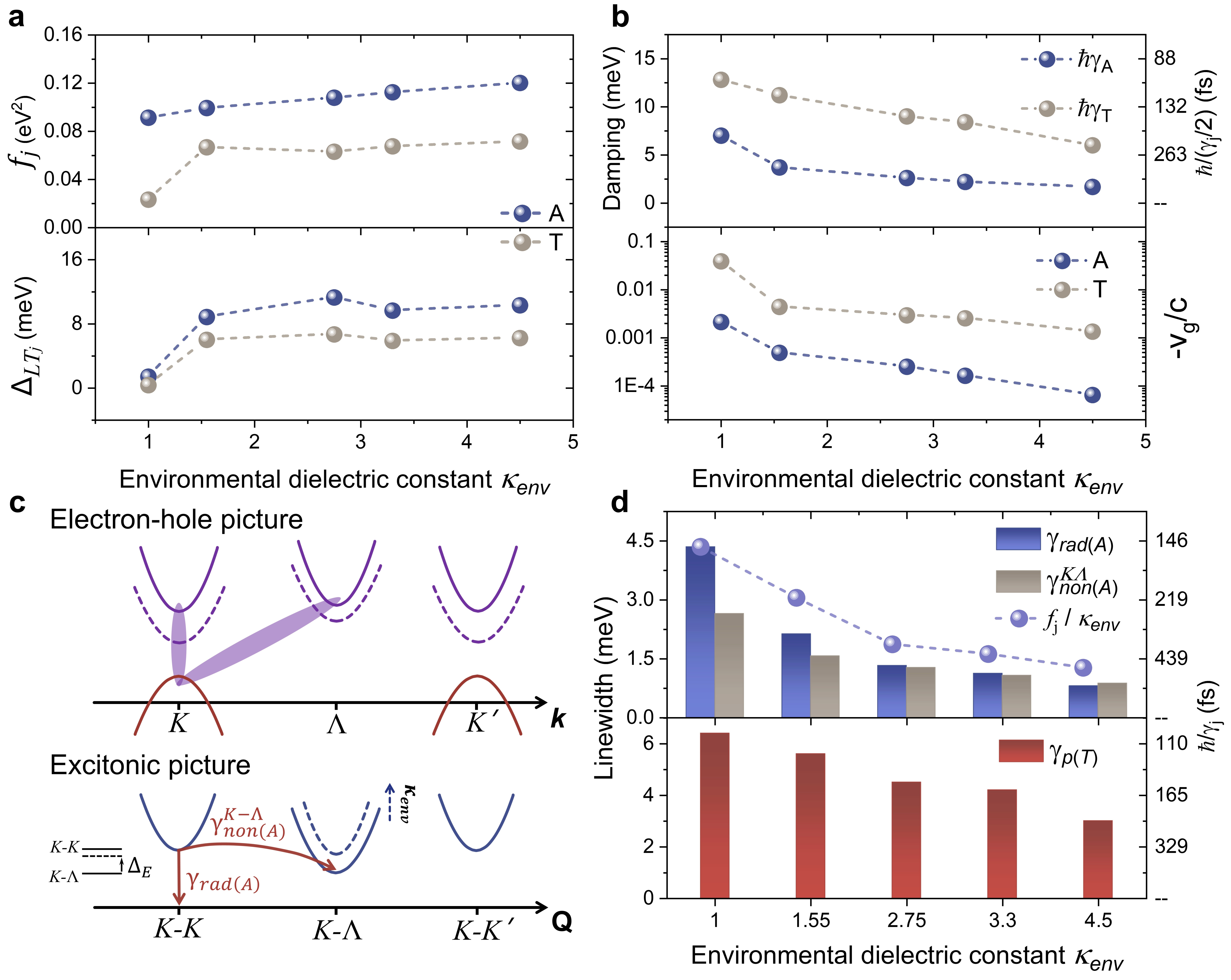}
\caption{\label{fig:wide}\textbf a Dependence of the oscillator strength (upper panel) and the longitudinal-transverse splitting energy (lower panel) of the ETPs on the environmental dielectric constant. \textbf b Dependence of the damping parameters (upper panel) and the ratio of the negative group velocity to the speed of light (-$\upsilon_g$/$c$, lower panel) of the ETPs on the environmental dielectric constant. \textbf c Schematic representation of $K$($K$-$K$), $\Lambda$($K$-$\Lambda$) and $K’$($K$-$K’$) dispersion in the electron-hole picture (upper panel) and exciton picture (lower panel). The dashed line stands for the exciton dispersions with increased dielectric screening. The $K$-$\Lambda$ excitonic transition energy in the exciton picture increases relative to $K$-$K$ exciton, with the $K$-$K$ transition energy fixed. \textbf d Calculated relaxation rate, $\gamma_{rad_{(A)}}$ and $\gamma_{non_{(A)}}^{K\Lambda}$, of the A exciton states (upper panel) and pure dephasing rate $\gamma_{p_{(T)}}$ of the trion states (lower panel) with respect to the environmental dielectric constant. The mauve dot-dashed line represents the ratios of $f_{(A)}/\kappa_{env}$, normalized to $\gamma_{rad_{(A)}}$ of the suspended WS$_2$ sample.}
\end{figure*}

\subsection{\label{sec:level3}DIELECTRIC SCREENING EFFECT ON EXCITON-TRION POLARTIONS}
The fitting curves of the five samples with different dielectric screenings are shown in the upper panels of Figures 2c-g. The values of various parameters adopted in the calculations are tabulated in TABLE I. The theoretical fitting curves have good consistency with the experimental reflectance curves. The lower panels of Figures 2c-g show the dispersion curves of the coupled A/T exciton-polaritons. For the calculated dispersion with $\gamma$=0 (light blue), the energy-momentum dispersion consists of three bands, consistent with the previous reports \cite{berz1991exchange,rana2021exciton}. The lower polariton branch of the trion states (LPB$_T$) begins with a photon-like line in the low wavevector region and then bends over to the exciton-like parabolic dispersion, approaching their transverse excitonic curves, ${\omega^2=\omega_{T_{(Tr)}}^2+\beta_{(T)}k^2}$, asymptotically. The upper polariton branch of the A exciton states (UPB$_A$) exhibits a transition from the exciton-like to photon-like dispersion with the increasing wavevector. In the coupled A/T exciton-polariton dispersion, the emergence of an exciton-like mixed polariton branch (MPB) represents a strong Coulomb coupling between the A excitons and the trions \cite{lagois1977dielectric}. The exciton-trion-photon coupling can be clearly seen in our results, as evidenced by the quantum-mechanical non-crossing rule for interacting energy levels \cite{rana2021exciton,36}. For the polariton dispersion with the non-zero damping parameter ($\gamma\neq0$, dark blue line in the lower panels of Figures 2c-g), the enhanced Coulomb interaction of the A and T oscillators can be further confirmed by the lineshape of the polaritonic dispersion relation. The Coulomb interactions produce the clear double resonance structures observed in the experimental reflectance spectrum near the region of $\omega_{T_{(Tr)}}\leq\omega\leq\omega_{L_A}$. The increased environmental screening results in clearer and sharper double resonance features in the polaritonic dispersion with $\gamma\neq0$. As the photon energy goes approaching $\omega_{T_{(Tr)}}$, the dispersion shows a photon- to exciton-like transformation. It is known that oscillator damping can induce a negative group velocity $\upsilon_g$=$d\omega/dk$, producing propagation suppression of the light-matter coupled polarization wave due to polariton-phonons scattering \cite{36}. Subsequently, the incident light will couple with A exciton states, and then the light-matter wave propagates in the form of exciton-like behavior with a positive group velocity while approaching $\omega_{L_A}$. Especially, strong dielectric screening induces faster dispersion transformation, reflecting weaker polariton-phonons scattering. With increasing the surrounding dielectric constant, the resonance energies of the A excitons and the trions approach each other, the Coulomb interactions between them thus tends to increase, eventually resulting in a stronger ETP effect \cite{rana2021exciton,lagois1977dielectric}. 

Further evidence for the dielectric screening-enhanced ETP effects can be obtained by analyzing the oscillator strength parameter and longitudinal-transverse branch splitting energy. Generally, for one resonance (or uncoupled) oscillator, the weaker dielectric screening will compress excitonic wave function spatially, thereby increasing excitonic oscillator strength \cite{wu2019exciton}. However, with reduced dielectric screening, the obtained oscillator strengths of the coupled A exciton and trion states become weaker in the five configuration samples, as evidenced in Figure 3a. Moreover, the longitudinal energy refers to the exciton-photon coupling energy at $\vec{k}$=0, whereas the transverse energy describes the zero exciton-photon interaction energy \cite{pantke1993damping}. There usually exists a finite splitting energy $\Delta_{LT}$ between the longitudinal and transverse branches, which is necessarily connected with a finite oscillator strength $f_j$. $\Delta_{LT}$ gives a measure of the interaction of excitons with photons, following the relationship of $\Delta_{LT_{(j)}}$=$E_{L_{(j)}}$-$E_{T_{(j)}}$=$f_j/(2\varepsilon_b E_{T_{(j)}})$ \cite{andreani1990exchange}. Strong environmental screening results in an increase trend of $\Delta_{LT}$ of both A exciton and trion, giving evidence for the enhancement of the ETP interactions. Furthermore, the contact with graphene can reduce the electron density of TMDC monolayers, leading to the suppression of the Coulomb coupling between A exciton and trion, that is, the uncoupled resonance \cite{rana2021exciton,froehlicher2018charge}. The reflectance amplitude of the single resonance of A exciton of monolayer ReSe$_2$ was observed to reduce as the dielectric screening it suffered was gradually enhanced \cite{qiu2019giant}. In fact, this behavior is also observed in the single resonance of B excitons related to the higher spin-orbit split states in the studied samples in the present work (i.e., as shown in Figure 1d).

\subsection{\label{sec:level3}DIELECTRIC SCREENING EFFECT ON POLARITON DECOHERENCE}
The strong interaction between the A excitons and trions was well interpreted, and a unique mechanism behind the dielectric screening tunability is the dephasing of ETPs. $\gamma_j$ corresponds to the exciton-polariton homogeneous broadening by  $\gamma_j$=$2\hbar/T_{2_{(j)}}$=$\hbar/T_{1_{(j)}}$+$2\gamma_{p_{(j)}}$, where $T_{2_{(j)}}$ is the corresponding oscillator decoherence lifetime, $T_{1_{(j)}}$ is radiative and non-radiative relaxation lifetime, and $\gamma_{p_{(j)}}$ is the pure dephasing rate \cite{36,pantke1993damping}. The obtained $T_{2_{(A)}}$ and $T_{2_{(T)}}$ values are approximately on the order of hundreds of femtoseconds (the upper panel of Figure 3d), which are consistent with the values previously measured by the optical two-dimensional coherent spectroscopy technology \cite{hao2016coherent}. Small $T_{2_{(T)}}$ means that $\gamma_{p_{(T)}}$ is a dominant factor in the trion dephasing process because of the quite long recombination time $T_{1_{(T)}}$, however, $\gamma_{p_{(A)}}$ is negligibly small in the A exciton dephasing process \cite{hao2017trion}. Such different decoherence pathways between the A exciton and trion states may be because the trion states interact with A exciton-polaritons through Coulomb coupling rather than photons. For the A exciton-polariton part, the non-radiative decay contains two kinds of decays, i.e., intravalley and intervalley phonon scattering channels. The intravalley and $K$-$K'$ intervalley scattering, usually negligible at low temperatures \cite{selig2016excitonic, jin2014intrinsic}. Significantly, even at low temperatures, the A exciton-polaritons can be scattered into the energetically lower-lying intervalley $K$-$\Lambda$ excitonic states, located approximately 70 meV below the $K$-$K'$ exciton due to the larger electron effective mass at $\Lambda$ valley in monolayer WS$_2$, through the optical $\Gamma$ or zone-edge acoustic $\Lambda$ phonon emission ($\gamma_{non_{(A)}}^{K\Lambda}$ in Figure 3c) \cite{selig2016excitonic, ferreira2022signatures}. Now we attempt to analyze the modulation mechanism of dielectric screening on the decoherence process of A exciton-polariton quantitatively. The reflectance amplitude of the monolayer WS$_2$ can be described by the polaritonic Elliot formula \cite{kira2006many, glazov2014exciton, fitzgerald2022twist}: 

\begin{eqnarray}
R_A(\omega)=\frac{\gamma^2_{rad_{(A)}}}{(\omega-\omega_{0_{(A)}})^2+(\gamma_{rad_{(A)}}+\gamma_{non_{(A)}}^{K\Lambda})^2}
\end{eqnarray}

where $\omega_{0_{(A)}}$ is the excitonic resonance frequency. Hence, the $\gamma_{rad_{(A)}}$ and $\gamma_{non_{(A)}}^{K\Lambda}$ are directly extracted by $\gamma_A$=$\gamma_{rad_{(A)}}$+$\gamma_{non_{(A)}}^{K\Lambda}$ and the maximum reflectance amplitude value of $R(\omega_{0_{(A)}})$. As shown in Figure 3d, based on the relation of $\gamma_{rad_{(A)}}\propto f_{(A)}/\kappa_{env}$  (dot-dashed line), the dominant increase of $\kappa_{env}$ shall result in a reduced $\gamma_{rad_{(A)}}$ \cite{glazov2014exciton}. Furthermore, in term of $\gamma_{non_{(A)}}^{K\Lambda}$, the transition energy of the $K$-$\Lambda$ exciton exhibits more sensitive behavior to the change of dielectric environments than the $K$-$K$ exciton, resulting in a gradual energy approaching between the $K$-$\Lambda$ and $K$-$K$ excitons with increasing the surrounding screening \cite{zibouche2021gw,su2022dark}, as shown in Figure 3c. As a result, the $K$-$\Lambda$ excitonic relaxation induced by the phonon emission would exhibit a smaller efficiency, and $\gamma_{non_{(A)}}^{K\Lambda}$ remarkably decreases. In the highly-screened hBN/WS$_2$/hBN sample, $\gamma_{rad_{(A)}}$ and $\gamma_{non_{(A)}}^{K\Lambda}$ make small contributions to the homogeneous broadening, which are the important factors for the extremely narrow reflectance linewidth of the hBN encapsulated WS$_2$ \cite{cadiz2017excitonic}. The pure dephasing rate usually refers to the elastic scattering process without energy loss. The Coulomb interactions between the A excitons and the trions, enhanced by the reducing dielectric screening, could potentially diminish the pure dephasing rate $\gamma_{p_{(T)}}$, as shown in the lower panel in Figure 3d. The reduced scattering of the ETPs is also reflected in the negative $\upsilon_g$/$c$ of the polaritonic dispersion relation, i.e., as seen in Figure 3b.

{\section{\label{sec:level4}CONCLUSIONS}}
In summary, we present an experimental demonstration of the existence of robust ETPs in a series of monolayer WS$_2$ samples with varied dielectric environments. Moreover, it is firmly shown that the Coulomb interactions between the exciton and the trion in the ETPs can be tuned by changing the environmental dielectronic constant. In addition, the distinct influencing mechanisms of the environmental dielectric screening on the decoherence process of the ETPs are unraveled. The demonstration of the existence of the robust ETPs, especially their tunability in 2D TMDC semiconductors may open intriguing scenarios of manipulating composite quasi-particles consisting of excitons, trions, and photons. The work may also materialize further exploration in the novel phase singularity and matter condensation by artificial dielectric engineering, such as the creation of recently emerged moiré superlattices.

\begin{acknowledgments}
This work is financially supported by the Natural Science Foundation of China (Grant Numbers 12074324 and 11374247). K.W. and T.T. acknowledge support from the JSPS KAKENHI (Grant Numbers 21H05233 and 23H02052) and World Premier International Research Center Initiative (WPI), MEXT, Japan.
\end{acknowledgments}

\bibliography{Reference/apssamp} 

\begin{widetext}

\section{SUPPORTING INFORMATIONS}
\textbf{Sample preparation and characterization.} Monolayer WS$_2$ flakes were prepared by using a mechanical exfoliating method from a high-quality WS$_2$ bulk crystal (HQ Graphene company) and then transferred onto transparent polydimethylsiloxane (PDMS) stamps (Gel-park). The identification and verification of the prepared monolayer WS$_2$ flakes were carried out with optical microscopy, PL, and Raman spectroscopy in the ambient atmosphere at room temperature. The suspended WS$_2$ samples were fabricated by transferring the PDMS/WS$_2$ onto the SiO$_2$/Si substrate with 10 $\mu$m grooves pre-patterned by reactive ion etching procedure. To obtain the hBN-supporting WS$_2$ samples, few-layer hBN flakes were transferred onto a SiO$_2$/Si substrate by the PDMS dry transfer method, and then the PDMS/WS$_2$ were attached to the target hBN flakes. And the hBN/WS$_2$/hBN samples were prepared by transferring few-layer hBN flakes onto the prepared hBN-supporting WS$_2$ samples. Note that before each transferring step, a thermal heating process at around 363 K for 5 minutes was usually conducted to enhance the adhesion force between the WS$_2$ monolayer and hBN or SiO$_2$/Si interface.

\textbf{Optical characterization.} Micro-PL and Raman measurements were achieved on a home-assembled multi-function-integrated micro-spectroscopic system equipped with a monochromator (Horiba iHR 550, 1200 grooves/mm grating and 2400 grooves/mm grating), a charge-coupled device (CCD) detector (Horiba Syncerity 7379, detection wavelength range of 200-1000 nm). A 532 nm continuous-wave laser was employed as the excitation source. The reflectance spectra were conducted by introducing a broadband picosecond laser (NKT, supercontinuum laser) into the above home-assembled system.
During the variable-temperature PL/Raman and reflectance spectroscopic measurements, the samples were mounted on the cold finger of a He-closed cycle cryostat (Montana Instruments S100, varying temperature range of 3.2-330 K, background vacuum pressure of 10$^{-4}$ torr). The excitation lights were focused on the sample surface via a ×100 objective lens and the light spot was about 1 $\mu$m. The reflectance spectra were given by $\Delta R$/$R$=($R_{sample}$-$R_{substrate}$)/$R_{substrate}$. The excitation light power was kept at 1 $\mu$W to minimize the possible unwanted influence of the excitation light in the micro-spectroscopic measurements. 

\renewcommand{\thefigure}{S1}
\begin{figure}
\includegraphics[width=0.8\textwidth]{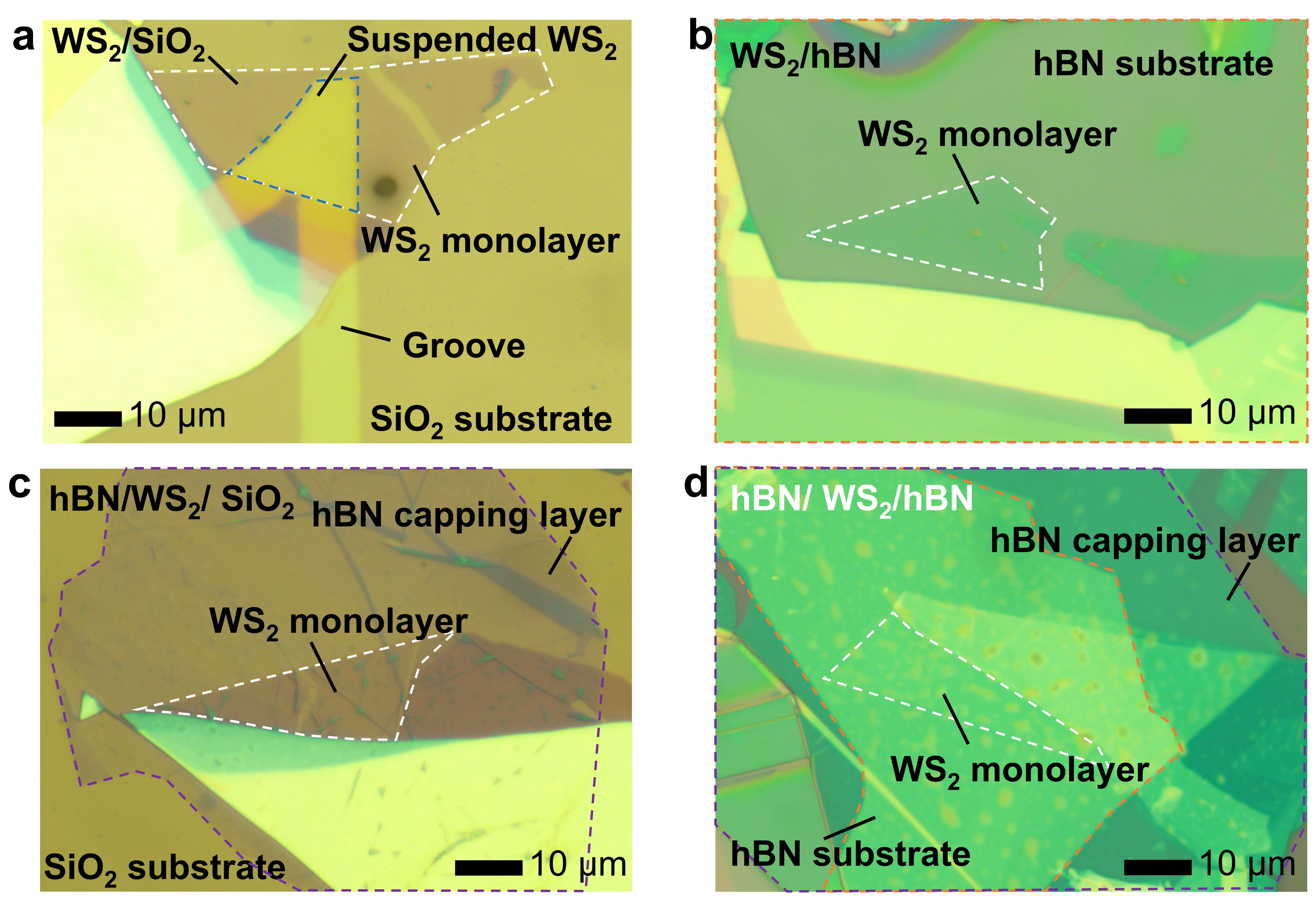}
\caption{\label{fig:wide} Optical images of the suspended WS$_2$ and WS$_2$/SiO$_2$ \textbf{(a)}, WS$_2$/hBN \textbf{(b)}, hBN/WS$_2$/SiO$_2$ \textbf{(c)} and hBN/WS$_2$/hBN \textbf{(d)}. The dashed white, orange and purple lines show the outline of monolayer WS$_2$, hBN bottom layers and hBN capping layers, respectively. The yellow number “1” and the dashed blue line parts refer to the 10 $\mu$m width groove and the suspended WS$_2$ in Figure S1 \textbf{(a)}, respectively. }
\end{figure}

\renewcommand{\thefigure}{S2}
\begin{figure}
\includegraphics[width=0.8\textwidth]{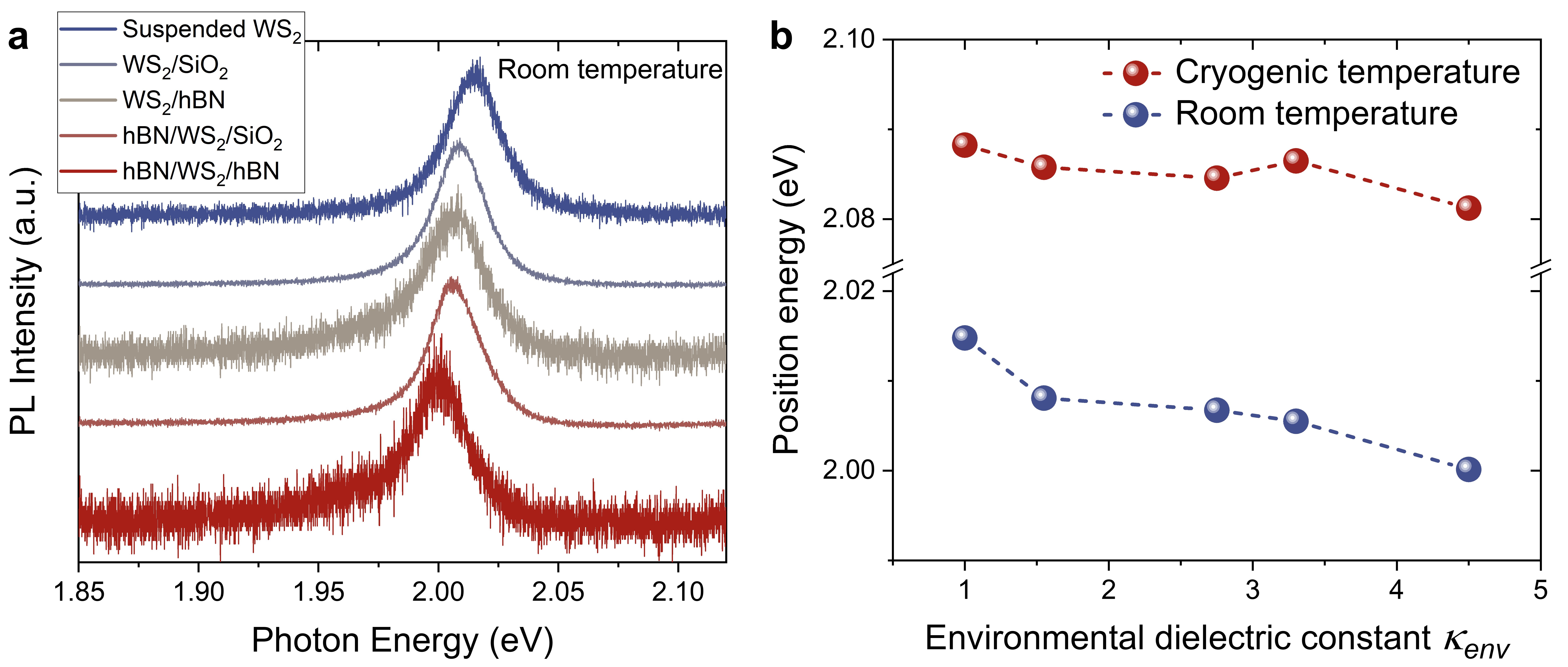}
\caption{\label{fig:wide}\textbf{a} Room-temperature PL spectra of the suspended WS$_2$, WS$_2$/SiO$_2$, WS$_2$/hBN, hBN/WS$_2$/SiO$_2$ and hBN/WS$_2$/hBN samples under the excitations of a weak 532 nm laser. \textbf{b} Dependence of the peak positions (solid circles) of the A exciton PL measured at cryogenic and room temperatures, respectively, on the environmental dielectric constant. }
\end{figure}

\renewcommand{\thefigure}{S3}
\begin{figure}
\includegraphics[width=0.8\textwidth]{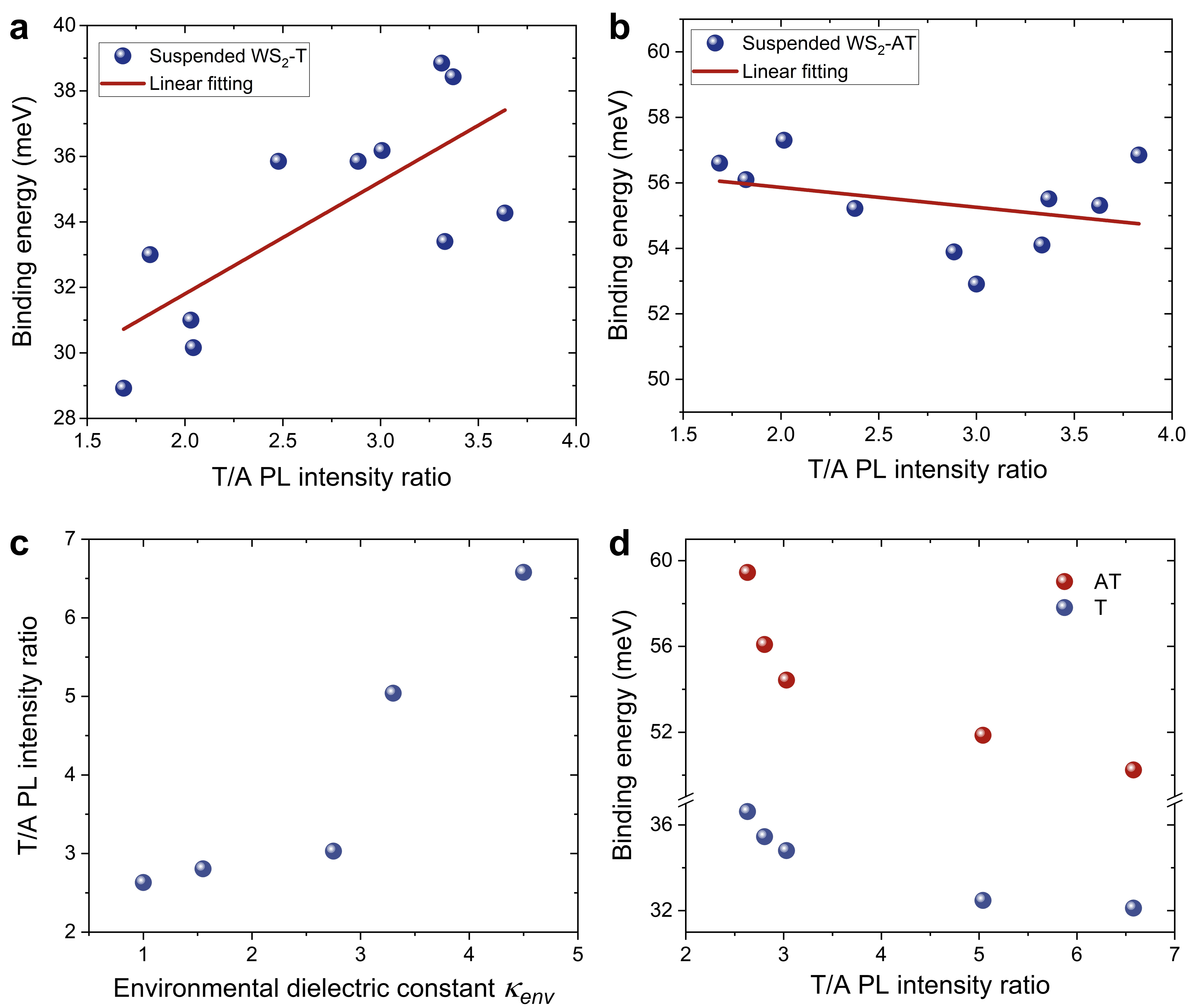}
\caption{\label{fig:wide} Binding energies (solid circles) of the T trions \textbf{(a)}  and the AT excitons \textbf{(b)}  vs. the T/A intensity ratio, extracted from the measured PL spectra of several suspended WS$_2$ monolayer samples. The solid lines represent linear fitting curves. \textbf{c}  T/A PL intensity ratio of the samples with different dielectric screenings. \textbf{d}  Binding energies of the T trions and the AT excitons vs. the T/A PL intensity ratio for the different samples with various dielectric screenings. }
\end{figure}

\end{widetext}

\end{document}